\documentclass[apj]{emulateapj}
\usepackage{graphicx}
\usepackage{epstopdf}
\usepackage{soul}
\usepackage{amsmath}
\usepackage{natbib}
\usepackage{color}

\def\lax{{$\mathrel{\hbox{\rlap{\hbox{\lower4pt\hbox{$\sim$}}}\hbox{$<$}}}$}}
\def\gax{{$\mathrel{\hbox{\rlap{\hbox{\lower4pt\hbox{$\sim$}}}\hbox{$>$}}}$}}

\slugcomment{Accepted for published in {\it The Astrophysical Journal}}

\begin{document}

\title{Detection of Prominent Stellar Disks in the Progenitors of
  Present-day Massive Elliptical Galaxies}
\author{Roozbeh H. Davari\altaffilmark{1,2},
Luis C. Ho\altaffilmark{3,4}, 
Bahram Mobasher\altaffilmark{1}, and 
Gabriela Canalizo\altaffilmark{1} 
}
\altaffiltext{1}{University of California, Riverside 900 University Avenue, Riverside, CA 92521, USA}
\altaffiltext{2}{The Observatories of the Carnegie Institution for Science 813 Santa Barbara Street, Pasadena, CA 91101, USA}
\altaffiltext{3}{Kavli Institute for Astronomy and Astrophysics, Peking University, Beijing 100871, P. R. China}
\altaffiltext{4}{Department of Astronomy, School of Physics, Peking University, Beijing 100871, P. R. China}

\begin{abstract}

Massive galaxies at higher redshifts ($\emph{z}$ $>$ 2) show different
characteristics from their local counterparts: They are compact
and most likely have a disk.  In this study, we trace the evolution of local
massive galaxies by performing a detailed morphological analysis,
namely, fitting single S\'{e}rsic profiles and performing bulge+disk
decompositions. We analyze $\sim$ 250 massive galaxies selected from
all CANDELS  fields (COSMOS, UDS, EGS, GOODS-South and GOODS-North). We confirm
that both star-forming and quiescent galaxies increase their
sizes significantly from $\emph{z}$ $\approx$ 2.5 to the present day. The global S\'{e}rsic
index of quiescent galaxies increases over time (from $n$ $\approx$
2.5 to $n$ $>$ 4), while that of star-forming galaxies remains roughly constant ($n$ $\approx$
2.5). By decomposing galaxy profiles into bulge+disk components, we find that
 massive galaxies at high redshift have prominent stellar disks, which are
 also evident from visual inspection of the images.
 By $z$ $\approx$ 0.5, the majority of the disks disappear and massive 
 quiescent galaxies begin to resemble the local elliptical galaxies. 
 Star-forming galaxies have lower bulge-to-total ratios ($B/T$)
than their quiescent counterparts at each redshift bin. The bulges of star-forming 
and quiescent galaxies follow different evolutionary histories, while their disks evolve
 similarly. Based on our morphological analysis and previous
 cosmological simulations,  we argue that major mergers, along with
 minor mergers, have played a crucial role in the significant size
 increase of high-\emph{z}  galaxies and the destruction of their
 massive and large-scale disks.
\end{abstract}


\section{Introduction}

Several studies have shown that at \emph{z} $\approx$ 2 a considerable 
fraction of the massive galaxies (stellar mass $M_{\star}$ $\approx$ 10$^{11}$ 
$M_{\sun}$) are compact compared to their local counterparts 
(e.g., \citealt{Daddi05}; \citealt{Cimatti08}; \citealt{vanderWel08};
\citealt{vanDokkum08}; \citealt{Damjanov09}; 
\citealt{Hopkins09}; \citealt{Cassata10, Cassata11}; \citealt{Mancini10};
\citealt{Newman12}; \citealt{Szomoru12}; \citealt{Williams14}). The rarity of compact massive galaxies at the present 
time implies a considerable size increase in the last 10 billion years 
(\citealt{vanDokkum08}; \citealt{Trujillo09}; \citealt{Taylor10};  
\citealt{vanDokkum10b}; but see \citealt{Saracco10}; \citealt{Valentinuzzi10}; 
\citealt{Ichikawa12}; \citealt{Poggianti13}). Recent comprehensive simulations 
have found that the commonly used methods for measuring the sizes of these 
galaxies, such as fitting single-component S\'{e}rsic (1968) function, is 
reliable (e.g., \citealt{Mosleh13}; \citealt{Davari14}; Davari et al. 2016), 
despite the fact that, in many instances, their sizes are comparable to the 
scale of the {\it Hubble Space Telescope (HST)}\ point-spread function (PSF). 

The compactness of high-\emph{z} massive galaxies strongly suggests that their 
formation process involved strong dissipation on rapid timescales (e.g., 
\citealt{Naab07}).  This can be accomplished by gas-rich major mergers (e.g., 
Barnes \& Hernquist 1992), cold gas flows (Dekel et al. 2009), or some 
combination of the two.  In support of such a scenario, the central regions of 
local massive ellipticals, the likely descendants of high-$z$ compact, massive 
galaxies, are old and have a high $\alpha$/Fe abundance ratio 
(\citealt{Thomas05}).  This indicates an early episode of violent star 
formation, which would naturally accompany a gas-rich, dissipative formation 
event.  Although major mergers have long been thought to transform disky 
galaxies to bulge-dominated systems (\citealt{Toomre77}; \citealt{Barnes92}), 
more recent simulations show that this may not be always the case.  In fact, 
gas-rich major mergers can leave large-scale disks (\citealt{Robertson06}; 
\citealt{Hopkins09}) if the gas retains significant 
angular momentum during the merger (\citealt{Springel05}), especially 
those that have a high gas fraction (\citealt{Hopkins09}).

The study of \citet{Toft14} lends credence to this picture.  These authors 
show that massive, evolved, compact galaxies at $\emph{z}$ $\approx$ 2 --- the 
so-called red nuggets --- are the direct descendants of the submillimeter 
galaxies (SMGs; Blain et al. 2002) at  $\emph{z}$ $>$ 3. SMGs are among the 
most luminous, rapidly star-forming galaxies known, with luminosities greater 
than 10$^{12}$ $L_{\odot}$ and star formation rates of $\sim 10^{2}-10^{3}$ 
$M_{\odot}$~yr$^{-1}$ (e.g., \citealt{Kovacs06}; 
\citealt{Magnelli10}; \citealt{Michalowski12}). Indeed, \citet{Toft14} show 
that the mass-size distribution and the mean stellar mass surface density of 
these two classes of high-redshift galaxies are similar. Both types are best 
fit by low S\'{e}rsic indices ($n$). Moreover, from a CO study of 30 local 
merger remnants, Ueda et al. (2014) find that the majority of the sources 
exhibit kinematic signatures of rotating molecular gas disks. Furthermore, 
\citet{Targett13} conclude that more than 95\% of SMGs have pure stellar disks 
or disk-dominated stellar structures; the distribution of axial ratios (their 
Figure 6) rejects the possibility that the sample is bulge-dominated.

The above arguments strongly suggest that high-$z$ massive galaxies should 
host large-scale stellar disks.  This hypothesis is attested by a number of 
studies. From the work of van der Wel et al. (2011), 50\% of massive galaxies 
at \emph{z} $>$ 2 are disk-dominated. Similarly, \citet{Chang13} find that 
massive galaxies at \emph{z} $>$ 1 have higher axial ratios than their lower
redshift counterparts, broadly consistent with the tendency for galaxies to
become noticeably rounder between $z \approx 3$ and 0 \citep{Patel13}.

Now the question remains: how have the red nuggets, which most likely 
contained a significant disk component at $\emph{z}$ $\approx$ 2 turn, into 
local giant ellipticals like M87, which demonstrably do {\it not} have a disk? 
We aim to trace this morphological transition. We do so by performing detailed 
two-dimensional modeling of the optical light distribution of massive galaxies
within 0.5 $<$ $\emph{z}$ $<$ 2.5. Besides fitting a traditional, simple 
single-component S\'{e}rsic function, when possible, we perform a bulge+disk 
decomposition of these massive systems.  Examining separately the bulge and 
disk structural properties, plus the luminosity bulge-to-total ratio ($B/T$), 
provides key indicators that can be missed by studying potentially 
multiple-component galaxies as a single system. For instance, from the
comprehensive morphological analysis by \citet{Bruce14b}, massive galaxies 
appear to transit from disk-dominated to bulge-dominated between $\emph{z} 
\approx 3$ and 1, with elliptical-like systems emerge at lower redshifts.
The bulge+disk decomposition carried out by Bruce et al. was done by fixing 
the S\'{e}rsic index of the bulge to $n=4$ and of the disk to $n=1$. In other 
words, all bulges were assymed to follow a de Vaucouleurs (1948) light 
profile. The simulations of Davari et al. (2016) show that this method can 
lead to biases in measuring the properties of the bulge and disk, depending 
on the size, $S/N$, and redshift of the
galaxy. For instance, fixing bulge $n$ can
  overestimate/underestimate the bulge/disk total brightness, and in
  general, the uncertainties tend to be greater when the bulge $n$ is
  fixed. Besides, by fixing the bulge 
S\'{e}rsic index, one cannot tell how the bulge density and shape evolve, and 
important information is lost.  Our study relaxes the restriction on the 
bulge profile shape, which results in more robust and
informative bulge+disk decompositions (Davari et al. 2016). 

The Cosmic Assembly Near-infrared Deep Extragalactic Legacy Survey 
(CANDELS;\citealt{Grogin11}; \citealt{Koekemoer11})\footnote{\tt 
http://candels.ucolick.org/}, provides an unprecedented chance to investidagate 
the morphological evolution of galaxies. In fact, one of the original science 
goals of CANDELS is to trace the bulge and disk growth in rest-frame optical 
wavelengths at 1 $<$ $z$ $<$ 3 (\citealt{Grogin11}). We take advantage of all 
wide and deep images taken in five well-known, widely separated fields: 
GOODS-South and GOODS-North (The Great Observatories Origins Deep Survey;
\citealt{GOODS}), UDS (UKIDSS Ultra-Deep Survey; \citealt{UKIDSS}), COSMOS 
(The Cosmic Evolution Survey; \citealt{COSMOSa}; \citealt{COSMOSb}), and EGS  
(The Extended Groth Strip; \citealt{EGS}). These collectively yield a 
statistically uniform and robust sample that mitigates cosmic variance.

The most massive galaxies in the local Universe are almost all quiscent 
(\citealt{Baldry12}), which is not the case at earlier epochs 
(\citealt{Whitaker11}). This means that massive star-forming galaxies have all 
quenched over time.  Quantifying the evolution of both quiescent and 
star-forming galaxies helps trace back the formation of massive ellipticals 
and understand the bigger picture.

We address three key questions:

1) How does the size and the shape of the light distribution (S\'{e}rsic 
index) of star-forming and quiescent massive galaxies evolve?

2) Do high-redshift massive galaxies have a prominent stellar disk? If yes, do 
their relative bulge fraction evolve significantly over the last 10 billion 
years?

3) What does the observed evolution of bulges and disks teach us about the 
history of massive galaxies?

Our findings show that the massive galaxies were compact and indeed more
disk-dominated at higher redshifts and became more bulge-dominated over time, 
converging to the population of massive ellipticals by today.  Only major 
mergers can effectively destroy large-scale disks.  Thus, while minor mergers 
were largely responsible for the significant size increase of high-\emph{z} 
galaxies, our results underscore that major mergers also played an important 
role in the morphological transformation of massive galaxies.

This paper is organized as follows. Section 2 provides details of the
sample definition, whcih uses techniques described in Section 3. The 
morphological analysis is presented in Section 4. Section 5 discusses the
implications of our results, and a summary is given in Section 6. Throughout 
this study we adopt a standard cosmology ($\emph{H}_0$ = 71 ${\rm km^{-1}\,
  s^{-1}\, Mpc^{-1}}$, $\Omega_m$ = 0.27, and $\Omega_{\Lambda}$ = 0.73) 
and AB magnitudes.

\section{Sample Definition}

We utilize CANDELS images and catalogs. Besides their high-quality near-IR 
photometry taken with {\it HST}/WFC3, the observations are complemented with 
deep {\it HST}/ACS optical images, mid-IR photometry from {\it Spitzer}, 
and near-UV observations from the ground.  This provides a reliable dataset 
for the determination of photometric redshifts and stellar masses.

The photometric redshifts are computed by combining 11 independent 
measurements (\citealt{Dahlen13}), each using different combinations of 
photometric redshift code, template spectral energy distributions, and priors. 

The median fraction difference between the photometric and spectroscopic 
redshifts is less than 0.01, with an rms scatter of $\sim0.03$ 
(\citealt{Dahlen13}).  As this study is mostly concerned with broad evolutionary 
trends between $\emph{z}$ $\approx$ 2.5 to 0.5, precise redshifts for 
individual objects are not essential to our analysis.

The final quoted stellar mass is the median of estimates from 10 different 
CANDELS teams, who used the same photometry and redshifts estimates but 
different fitting codes, assumptions, priors, and parameter grid 
(\citealt{Mobasher15}; \citealt{Santini15}).  For massive galaxies, there is 
good agreement between CANDELS and 3D-HST (\citealt{Skelton14}; 
\citealt{Santini15}). \citet{Mobasher15} perform extensive simulations to
quantify the different sources of errors and uncertainties, using 10 ten 
independent methods and mock galaxy catalogs with a range of redshifts, 
masses, and spectral energy distributions. They concluded that different 
methods have comparable scatter of 0.136 dex, with no significant bias.

We employ the CANDELS $H$-band images and the accompanied catalogs to analyze
the evolution of the rest-frame optical properties of massive galaxies between 
$z \approx 2.5$ and 0.5.  The observed $H$ magnitudes of
  our sample range from 24 - 17 mag (measured by Single S\'{e}rsic fit),
  corresponding to the $V$-band rest-frame of 15.5 - 12
  mag. We estimated the rest-frame magnitudes by
    constructing SEDs of individual galaxies, shifting them to \emph{z}=0,
    and convolving the resulting rest-frame SEDs with the $V$-band
    response function.  
The high resolution (pixel scale = 0.06$^{\arcsec}$),
bright limiting magnitude (5 $\sigma$ $\approx$ 27 mag), and wide areal 
coverage of the CANDELS fields, coupled with the availability of physical 
parameters (photometric redshift, stellar mass) for individual galaxies, make
this dataset unique and ideal for our photometric analysis.

\begin{figure}[t]
  \centering
\includegraphics[width=75mm]{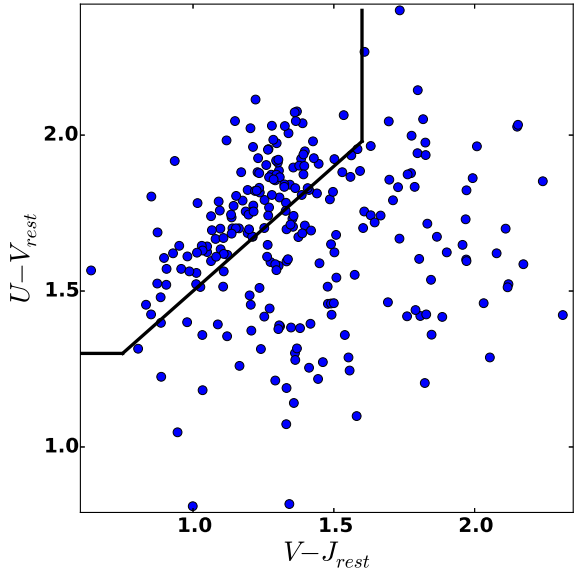}
  \caption{$UVJ$ color-color diagram is
    used for distinguishing quiscent galaxies from 
    star-forming galaxies. The quiescent galaxies populate the top
    left region of the diagram. \label{fig:UVJ}}
\end{figure}

\begin{figure}[t]
  \centering
\includegraphics[width=75mm]{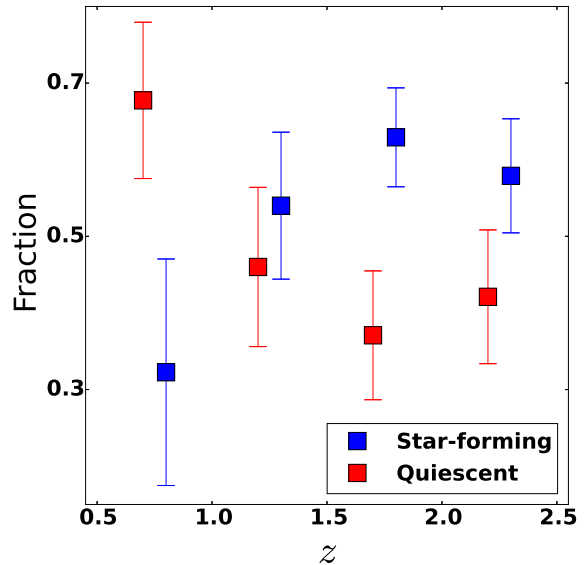}
  \caption{Fraction of massive star-forming and quiescent galaxies
    in redshift range 0.5 $<$ $\emph{z}$ $<$ 2.5. 
    Error bars show our sample proportions
    standard deviation. \label{fig:class_fraction}}
\end{figure}

We use the rest-frame $UVJ$ color-color diagram to separate quiscent galaxies 
from star-forming galaxies (see, e.g., \citealt{Labbe06}; \citealt{Wuyts07}; 
\citealt{Williams09}; \citealt{Patel13}).  We use the selection criteria of 
\citet{Patel13} to differentiate between these two types of galaxies (Figure 
\ref{fig:UVJ}). Quiescent galaxies populate a region defined by 

\begin{eqnarray}
U - V & > & 1.3 \\ \nonumber
V - J & < & 1.6 \\ \nonumber
U - V & > & 1.08(V - J) + 0.43,
\end{eqnarray}

\noindent where $U$, $V$, and $J$ are rest-frame magnitudes, calculated using 
the {\tt EAZY} photometric redshift code (\citealt{Brammer08}) and templates 
from \citet{Muzzin13}.  The fraction of star-forming and quiescent galaxies in 
our sample at different redshifts is shown in Figure \ref{fig:class_fraction} 
and Table 1. It can be seen that at $\emph{z}$ = 2.5 most massive galaxies 
were star-forming. By $\emph{z} \leq 1$, the majority of massive galaxies are
quenched, in agreement with the findings of \citet{Brammer11} and 
\citet{Patel13}.

\begin{figure}[t]
  \centering
\includegraphics[width=80mm]{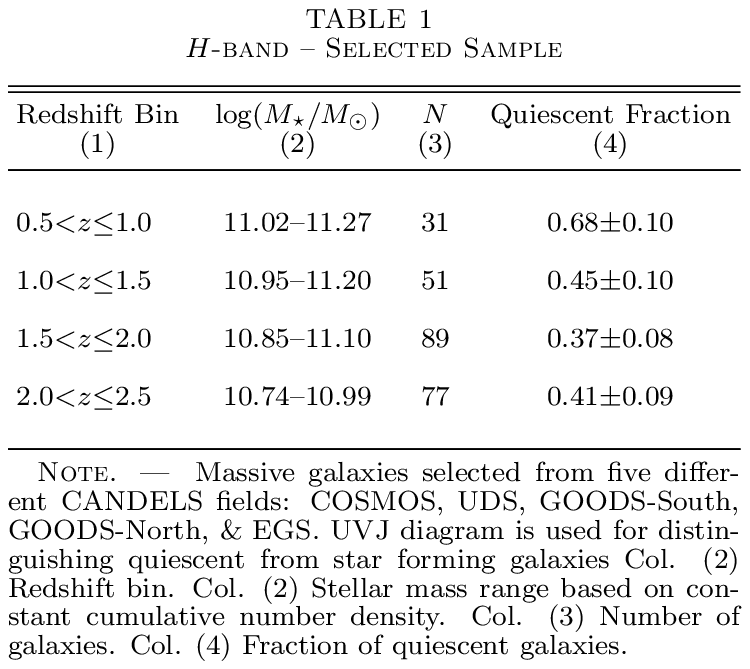}
\end{figure}

We choose massive galaxies based on number density selection rather than a 
fixed stellar mass limit.  For a chosen cumulative number density, we rank 
galaxies according to their stellar mass and chose galaxies of the same rank at 
different redshifts.  \citet{Mundy15}, using the Millennium Simulation results
(\citealt{Millennium}; \citealt{Lemson06}), show that the former is more 
reliable for tracing the true evolution of the average stellar mass below 
\emph{z} = 3. Number density selection, despite its limitations, is more 
physically motivated (\citealt{Leja13}). For instance, red nuggets have 
doubled their stellar masses in the last 10 billion years 
(\citealt{vanDokkum10b}). In other words, local massive galaxies were less 
massive in the past and could be left out of a fixed stellar mass selection. 
To trace the evolution of massive galaxies, it is more sensible to select
galaxies at a constant cumulative number density (\citealt{vanDokkum10b}; 
\citealt{Brammer11}; \citealt{Papovich11}; \citealt{Patel13}).  We use the 
criteria of Patel et al. (2013) for selecting galaxies at a fixed cumulative 
number density, $n_c$. Their Figure 2 shows that $n_c = 1.4\times10^{-4}$ 
Mpc$^{-3}$ corresponds to $M_{\star} = 10^{10.8}$ and $10^{11.1}\,M_\odot$ at 
$z = 2.5$ and 0.5, respectively.  Local ($z \approx 0$) galaxies with this 
corresponding stellar mass ($M_{\star} \approx 10^{11.2}\,M_\odot$)
are predominantly quiescent (\citealt{vanderWel09}; \citealt{Baldry12}) and 
have large axial ratios (\citealt{vanderWel09}), and therefore massive 
ellipticals.  

Our sample consists of $\sim$250 massive galaxies, whose properties are
summarized in Table 1. The mass range for each redshift bin takes into 
account systematic uncertainties in the stellar mass estimate.

\section{{\tt GALFIT} Modeling}

\begin{figure}[t]
  \centering
\includegraphics[width=142mm,angle=90]{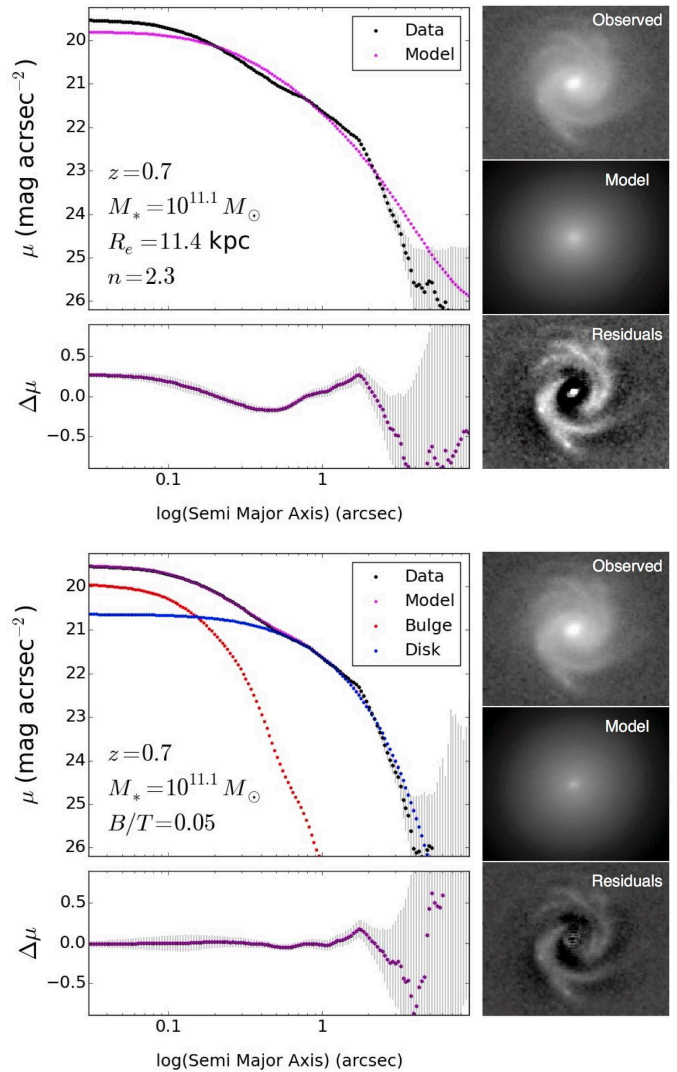}
  \caption{Diagnostic plots used to examine the goodness of a fit. Top
    left panels show the mean 
surface brightness ($\mu$) profile of the galaxy, the {\tt GALFIT} fit
model, and the bulge and disk components (in
cases of bulge+disk decomposition). Bottom
left panels show the residuals between the model and the observed mean
surface brightness. The error bars are calculated using
the RMS of the image background and the surface brightness measurement error
calculated by
{\tt ellipse} in {\tt IRAF}. The right panels, from top to bottom, show the observed galaxy, the
{\tt GALFIT} model, 
and the residuals, respectively. The top panel makes it clear that the fitted galaxy has a bulge-line central concentration and spiral
arms, and hence a disk; the model is trying to accommodate both components.
The bottom light
profile plots show that the bulge+disk decomposition reproduces the light
profile of the galaxy well. \label{fig:spiral_diagnostic}}
\end{figure}

\begin{figure}[t]
  \centering
\includegraphics[width=140mm,angle=90]{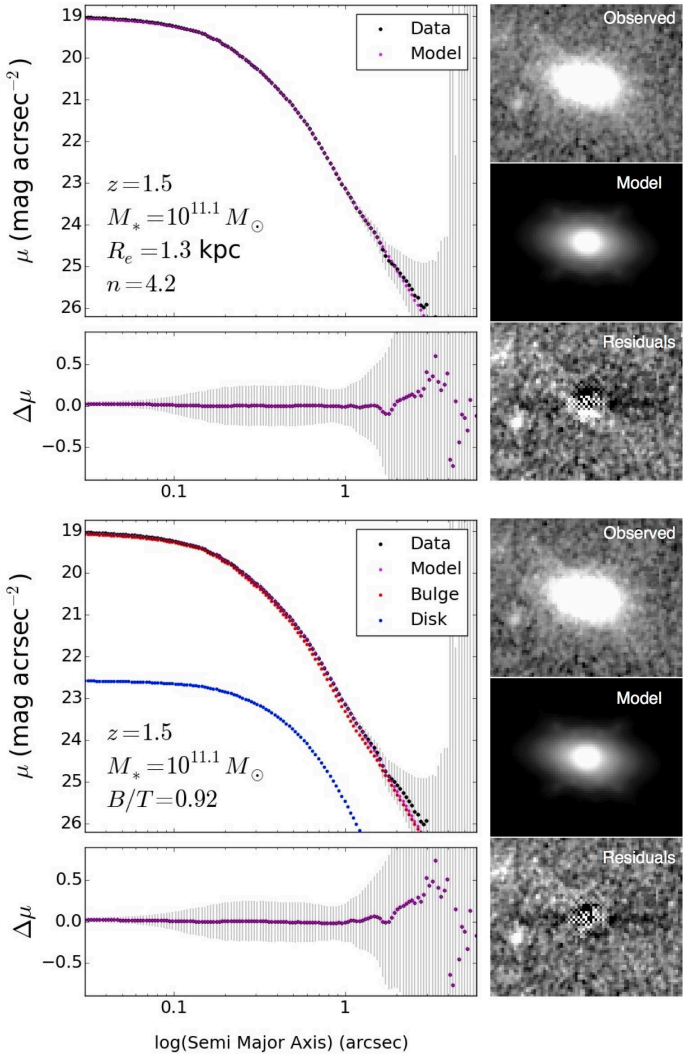}
  \caption{Similar to Figure \ref{fig:spiral_diagnostic}. This galaxy
is basically a single spheroidal, and adding a second component does not improve
the model fit significantly. \label{fig:elliptical_diagnostic}}
\end{figure}

We use {\tt GALFIT 3.0} (\citealt{Peng10}) as the main modeling tool. 
{\tt GALFIT} is a powerful, simple-to-use image analysis code to fit the light 
distribution of galaxies and other objects.  Of the several available options, 
we mainly use the S\'{e}rsic (1968) function to fit the surface brightness 
distribution: 

\begin{equation}
\Sigma(R) = \Sigma_e \ {\rm exp} {\left \{-\kappa \left [ \left (\frac{R}{R_e} \right )^{1/n} - \  1\right ]\right \}},
\label{eq:sersic}
\end{equation}

\noindent where $\Sigma_e$ stands for the surface brightness, $R_e$ is the 
half-light (effective) radius, $n$ is the S\'{e}rsic index that governs the 
shape of the profile, and $\kappa$ is a variable that depends on $n$ 
(\citealt{Ciotti91}).  The S\'ersic function is a generalization of the 
special cases of an exponential profile (\emph{n} = 1) used to model disks 
(\citealt{Freeman70}) and the \emph{$R^{1/4}$} law (\emph{n} = 4) traditionally
used to model elliptical galaxies and bulges (\citealt{deVaucouleur48}).  
Modern studies recognize that ellipticals and bulges have a more varied range 
of $n$ (e.g., Caon et al. 1993; Andredakis \& Sanders 1994; 
\citealt{Blanton03}; \citealt{Fisher08}).

Several inputs are needed to perform fit: a PSF model, a ``sigma'' image, and 
(sometimes) a bad pixel mask.  For each galaxy, we use the $H$-band CANDELS 
hybrid PSF corresponding to its field (e.g., UDS, COSMOS, etc.; 
\citealt{vanderWel12}). Hybrid PSFs are built by combining a stacked empirical 
stellar PSF and a synthetic {\tt TinyTim} (\citealt{Krist11}) PSF. CANDELS 
weight maps are used as the input sigma images.  As the field around each 
galaxy is chosen to be more than 10--15 times larger than the size of the 
galaxy, there are usually several other objects in the field that need to be 
masked.  As in Davari et al. (2014), we use 
{\tt SExtractor}\footnote{\tt http://www.astromatic.net/software/sextractor} 
(Bertin \& Arnouts 1996) to identify bright field objects and create a bad 
pixel mask that covers twice the area detected by {\tt SExtractor}.

For any given galaxy in the sample, our primary goal is to ascertain whether 
its light distribution, apart from a central bulge, shows evidence for an 
additional disk component, and if so, to determine its relative light fraction.
We model each galaxy twice, first with a single-component S\'ersic fit, and 
then with a two-component fit consisting of a bulge and a disk. The bulge is 
assigned a S\'ersic function with $n$ allowed to vary, and the disk fixed to 
an exponential.  We then carefully examine the residuals to determine the 
merits of the two models.

Depending on the complexity of the {\tt GALFIT} model, the initial parameters 
(guesses) can have a large effect on the fit. For single-component fits, 
unless the initial guesses are very far off the actual values, the initial
parameters do not have a major effect. Regardless, we use one-dimensional 
light profiles obtained by {\tt IRAF}/{\tt ellipse} (\citealt{Jedrzejewski87}) 
to obtain reliable initial guesses. We construct a curve-of-growth of the 
light distribution to estimate the effective radius, total luminosity, axial 
ratio, and position angle (for more details, see Davari et al. 2014).  
Appropriate initial guesses become much more important for the bulge+disk 
decompositions. For this type of modeling, we again use the one-dimensional 
light profile to obtain initial inputs. Assuming that the disk component 
follows an approximately exponential profile, we look for the part of 
the profile that traces a straight line in logarithmic space.  Depending on 
the $B/T$,  this region is located between 2 and 5 $R_e$, where the effect of 
the bulge is minimal. A straight line fitted to that section of the light 
distribution (in logarithmic space) provides an estimate for the disk scale 
length and central surface brightness. The total brightness obtained from a 
single S\'{e}rsic fit is used 
to find the total luminosity of the bulge component, and therefore $B/T$. Each 
galaxy is fit numerous times with different initial guesses for bulge $R_e$ 
and $n$, in addition to all the estimated parameters. While fitting a single 
S\'{e}rsic function might require only a few iterations, bulge+disk 
decompositions can require several fits with different initial guesses. The 
diagnostic plots (explained below) are imperative for evaluating the goodness 
of a fit. Lastly, the fitted sky component is left as a free parameter, and its 
initial values are set to zero. The simulations performed by \citet{Davari14} 
and Davari et al. (2016) show that once the field size of the image is more 
than 10 times larger than the galaxy, {\tt GALFIT} can measure the sky reliably.
 
Figures \ref{fig:spiral_diagnostic} and \ref{fig:elliptical_diagnostic} give 
examples of the diagnostic plots used to examine the goodness of a fit and 
whether or not a second component is needed.  The top left panels show the mean
one-dimensional surface brightness ($\mu$) profile of the galaxy, the final 
{\tt GALFIT} model, and the bulge and, if necessary, the disk components.  The 
bottom left panels show the residuals (model $-$ galaxy). The error bars are 
calculated using the rms of the image background and the galaxy flux 
measurement error (output from {\tt ellipse}). The right panels, from top to 
bottom, illustrate the two-dimensional image of the galaxy, model, and 
residuals.

The top right panel of Figure \ref{fig:spiral_diagnostic} clearly reveals 
that the galaxy, in addition to a bright central concentration, has a disk and 
spiral arms. The one-dimensional profile in the left panel confirms that the 
galaxy contains complex structure. The single-component model is trying to 
capture the most of the combination of two components, but the fit is clearly 
inadequate. The bulge+disk decomposition, by contrast, reproduces well the 
$\mu$ profile on the bottom left panel.  And not surprisingly, this galaxy is 
extremely disk-dominated: the best fit yields $B/T$ = 0.05.   At the other 
extreme, Figure \ref{fig:elliptical_diagnostic} showcases a galaxy that is
basically a just single big bulge; adding a second component does not improve
the fit significantly. The high $B/T$ of this galaxy (0.92) validates this 
hypothesis. 

\begin{figure}[t]
  \centering
\includegraphics[width=90mm]{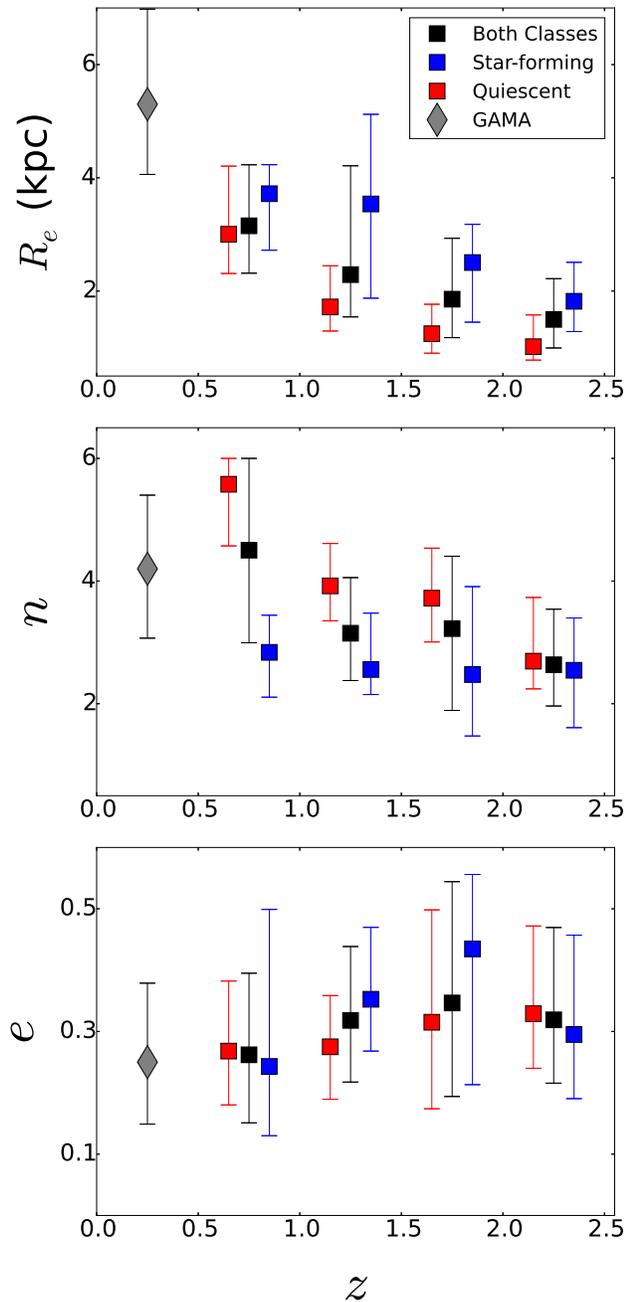}
  \caption{Results of single S\'{e}rsic fits.
Top, middle, and bottom panels show the redshift evolution of effective
radius ($R_e$), S\'{e}rsic index ($n$), and ellipticity ($e$).
Red, blue, and black filled boxes show the median in each redshift
bin for quiescent galaxies, star-forming galaxies, and both types combined. 
The gray filled diamond shows the median value for a
sample from the second data release of GAMA (\citealt{GAMA}; \citealt{GAMA2}) with mass range corresponding to our
number density selection criteria. Their morphological
parameters are derived from single S\'{e}rsic fits, consistent with our method.
The error bars correspond to the interquartile range of different measurements. 
\label{fig:singleSersic}}
\end{figure}

\begin{figure}[t]
  \centering
\includegraphics[width=80mm]{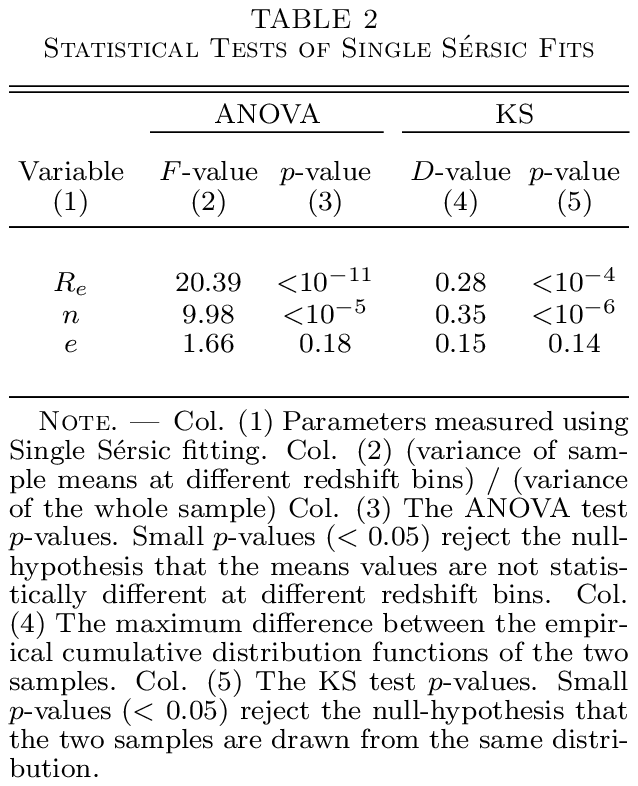}
\end{figure}

\begin{figure}[t]
  \centering
\includegraphics[width=80mm]{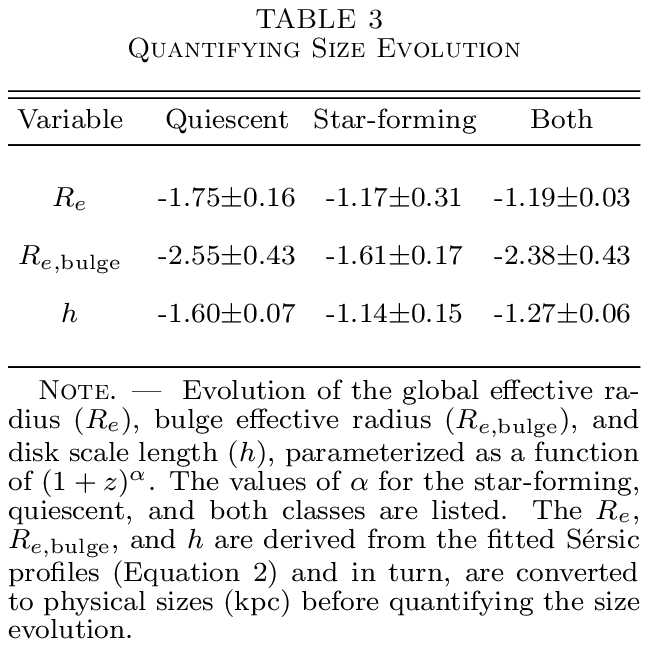}
\end{figure}

\begin{figure}[t]
  \centering
\includegraphics[width=90mm]{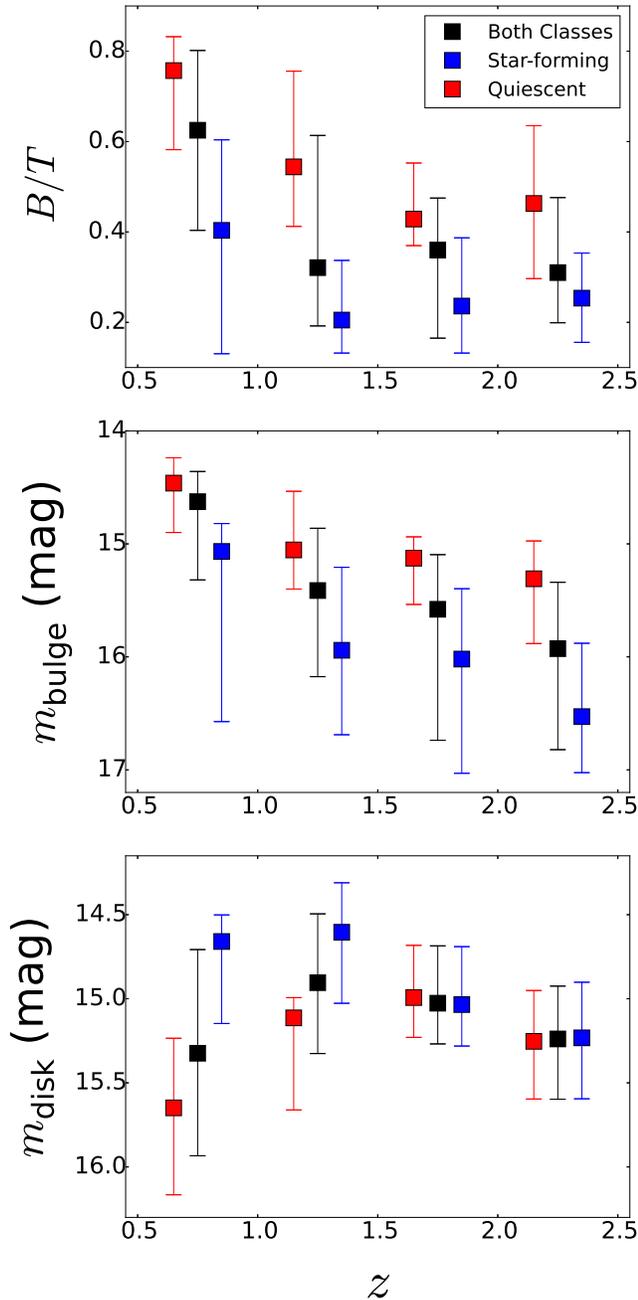}
  \caption{Results of bulge+disk decomposition.
Top, middle, and bottom panels show the redshift evolution of
flux bulge-to-total ratio ($B/T$), bulge magnitude ($m_{\rm bulge}$),
and disk magnitude ($m_{\rm disk}$).
The reported magnitudes are rest-frame magnitude in V-band.
Red, blue, and black filled boxes show the median in each redshift
bin for quiescent galaxies, star-forming galaxies, and both types combined. The error
bars correspond to the interquartile range of different
measurements. \label{fig:BT}}
\end{figure}

\begin{figure}[t]
  \centering
\includegraphics[width=90mm]{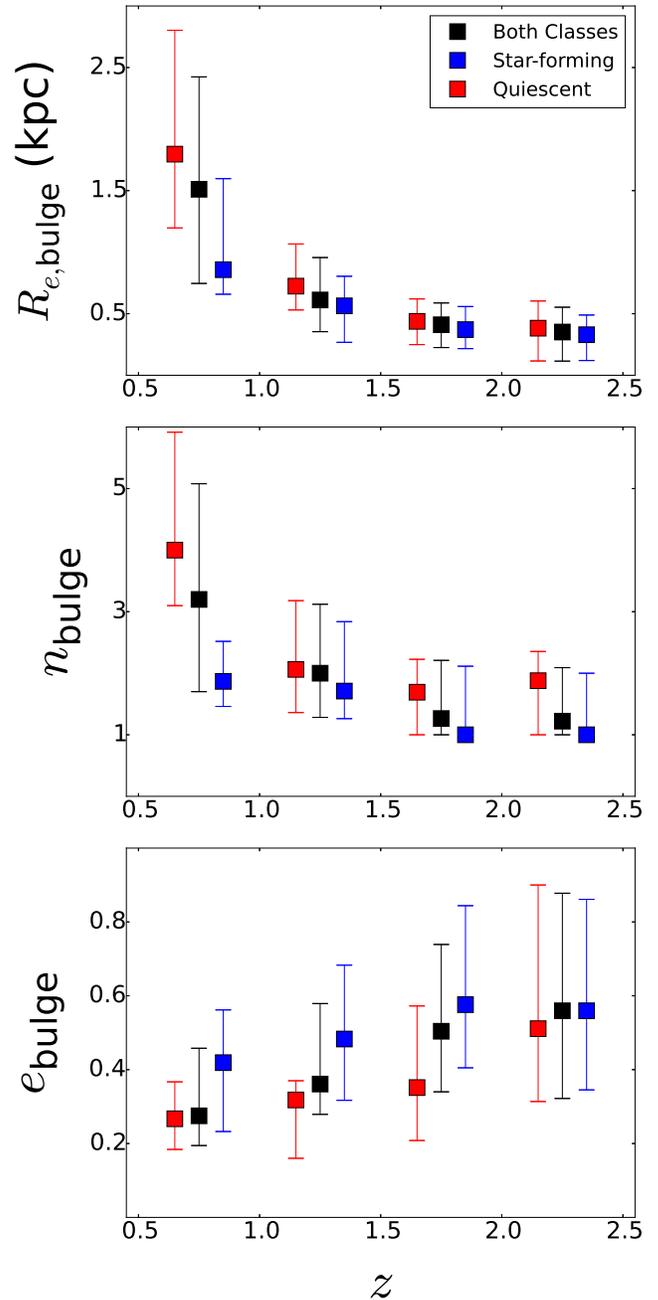}
  \caption{Results of bulge+disk decomposition.
Top, middle, and bottom panels show the redshift evolution of
bulge effective radius ($R_{e,{\rm bulge}}$), bulge S\'{e}rsic index
($n_{\rm bulge}$), and bulge ellipticity ($e_{\rm bulge}$).
Red, blue, and black filled boxes show the median in each redshift
bin for quiescent galaxies, star-forming galaxies, and both types combined. The error bars correspond to the interquartile range of different
measurements. \label{fig:Bulge}}
\end{figure}

\begin{figure}[t]
  \centering
\includegraphics[width=90mm]{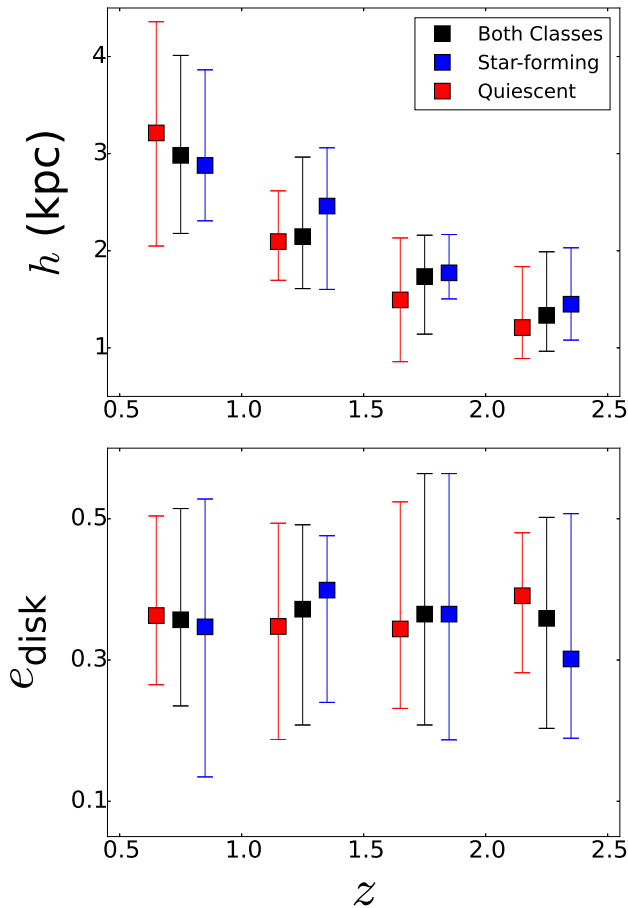}
  \caption{Results of bulge+disk decomposition.
Top and bottom panels show the redshift evolution of
disk scale length ($h$) and disk ellipticity ($e_{\rm disk}$).
Red, blue, and black filled boxes show the median in each redshift
bin for quiescent galaxies, star-forming galaxies, and both types combined. The error bars correspond to the interquartile range of different
measurements.\label{fig:Disk}}
\end{figure}

\begin{figure}[t]
  \centering
\includegraphics[width=80mm]{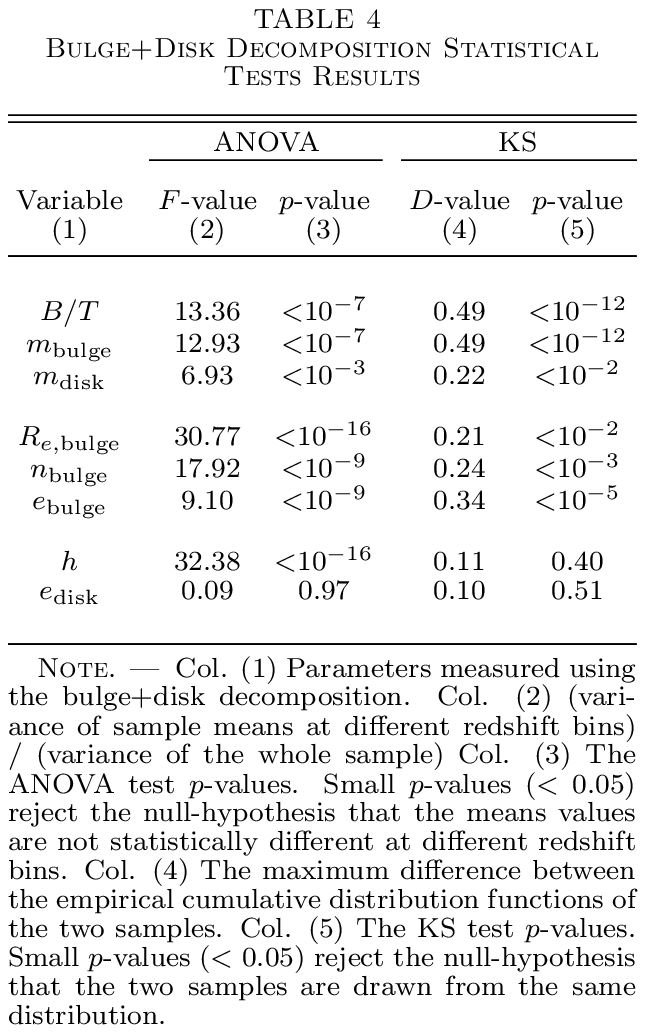}
\end{figure}

\begin{figure*}[t]
  \centering
\includegraphics[width=180mm]{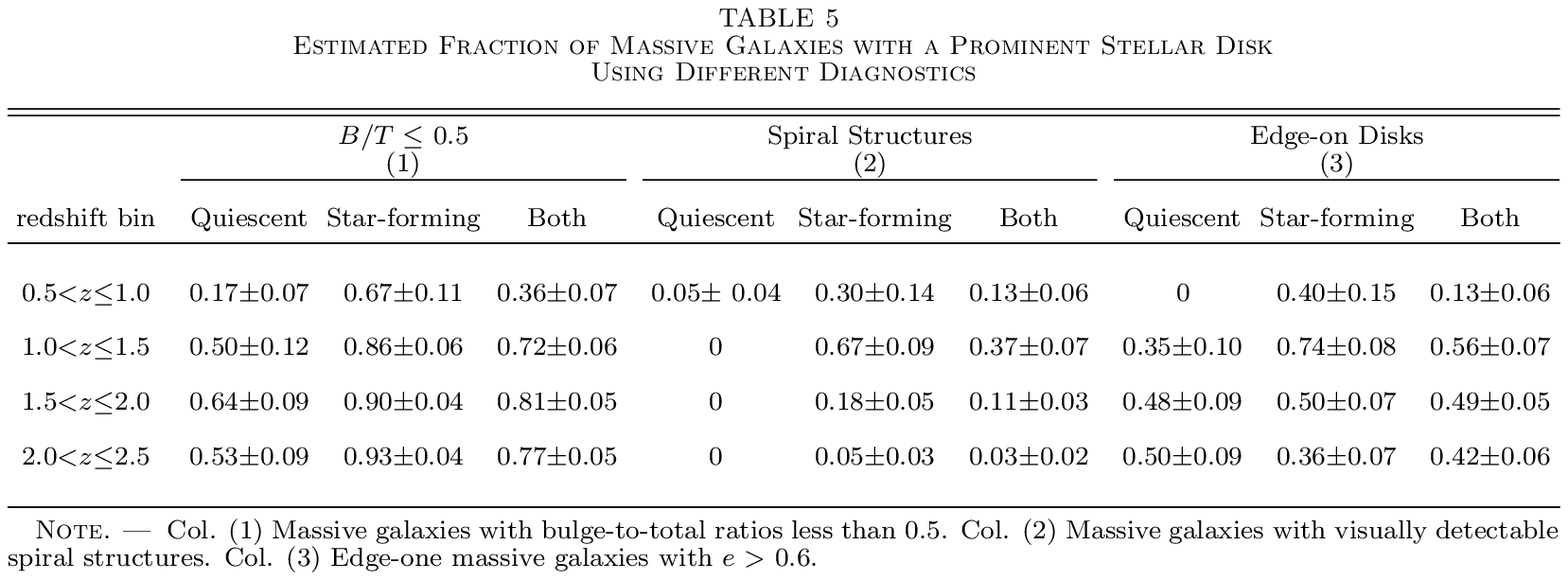}
\end{figure*}

\begin{figure*}[t]
  \centering
\includegraphics[width=180mm]{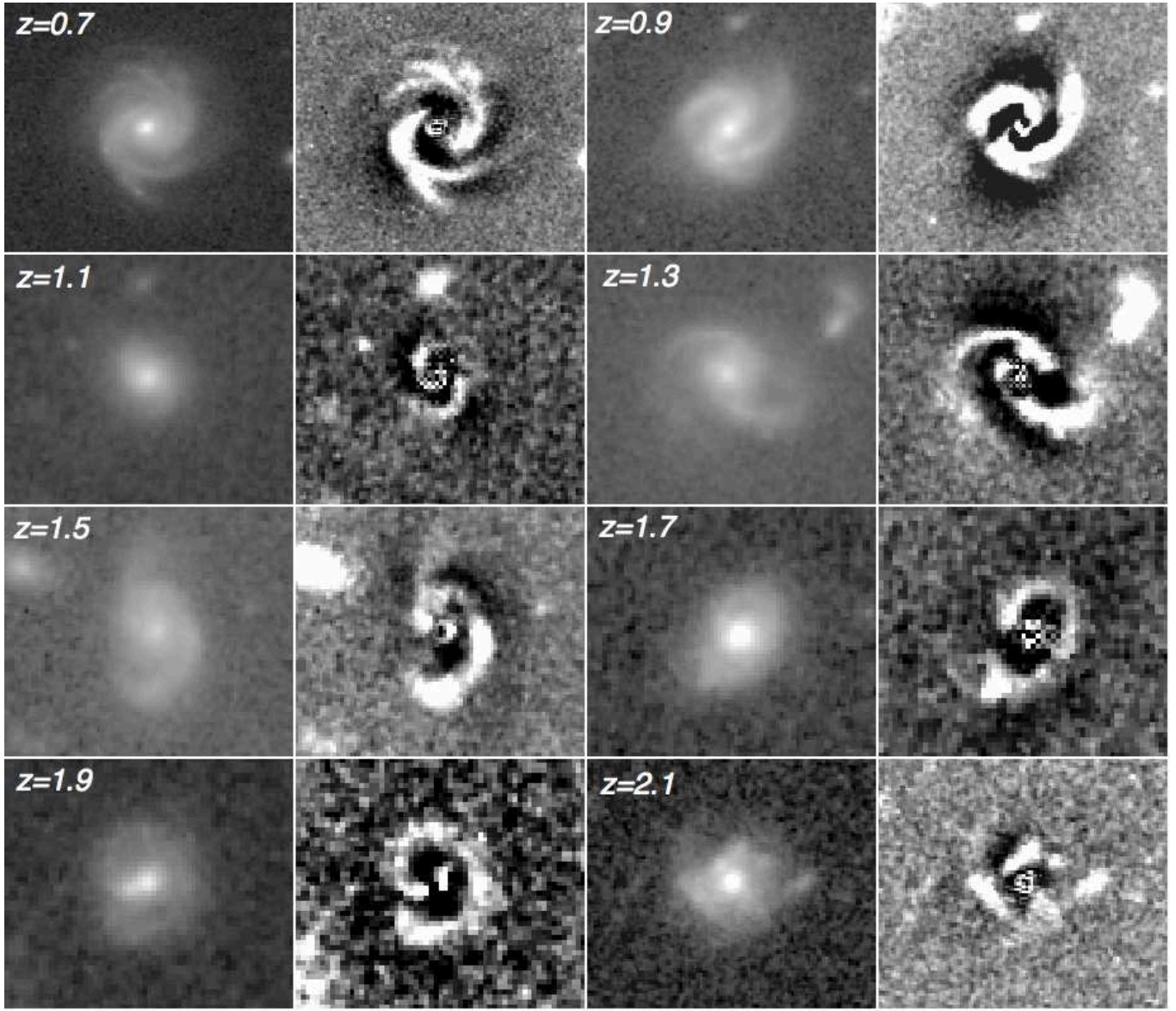}
  \caption{Examples of galaxies with apparent spiral structures
    at different redshifts. The residual images, 
    after removal of the bulge+disk model from the original galaxy
    image, allows for more effective detection of fine substructure.
    \label{fig:spiral_images}}
\end{figure*}

\begin{figure*}
  \centering
\includegraphics[width=180mm]{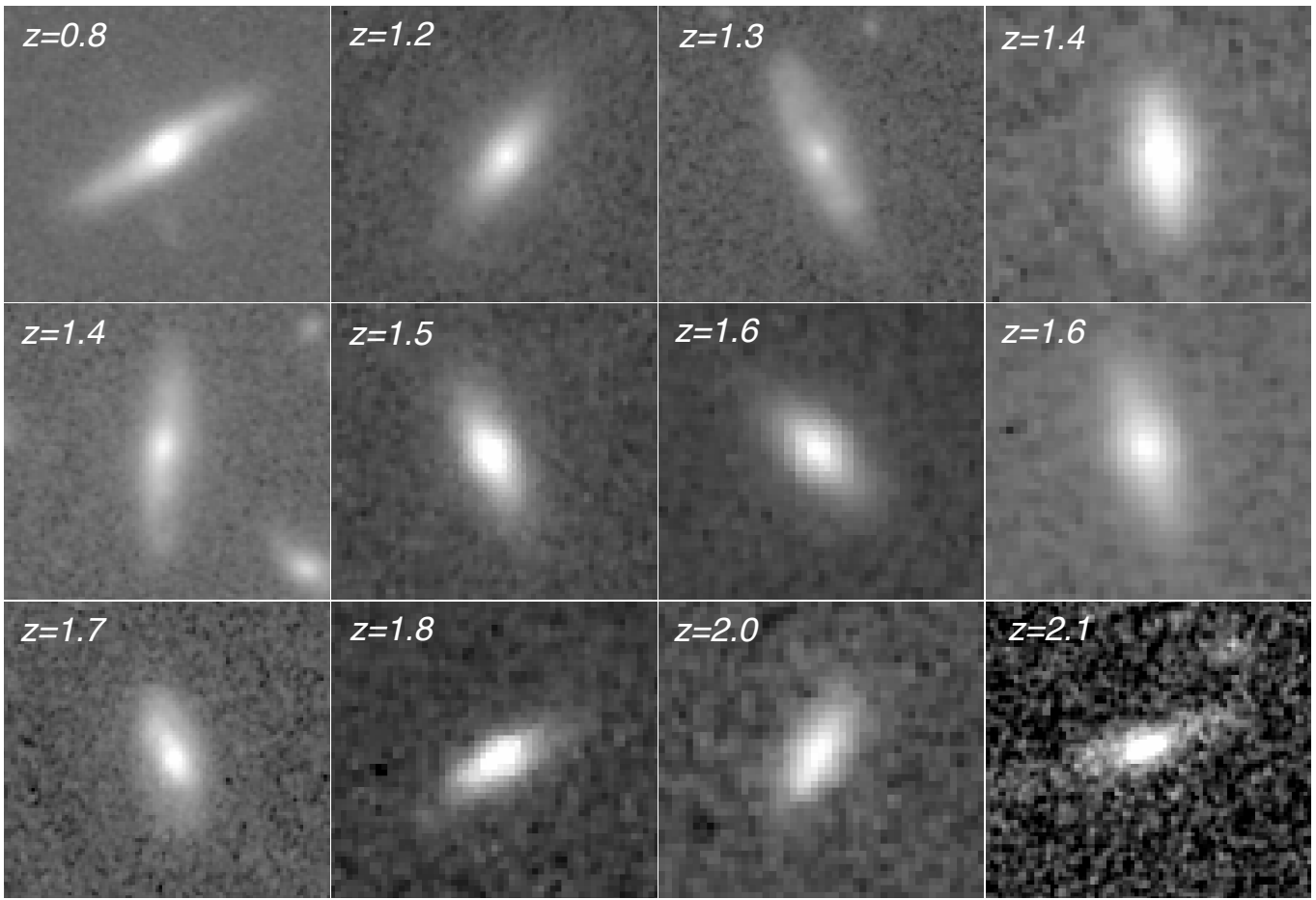}
  \caption{Examples of massive galaxies with an edge-pn disk at 
   different redshifts.\label{fig:thinDisk}}
\end{figure*}

\begin{figure*}[t]
  \centering
\includegraphics[width=185mm]{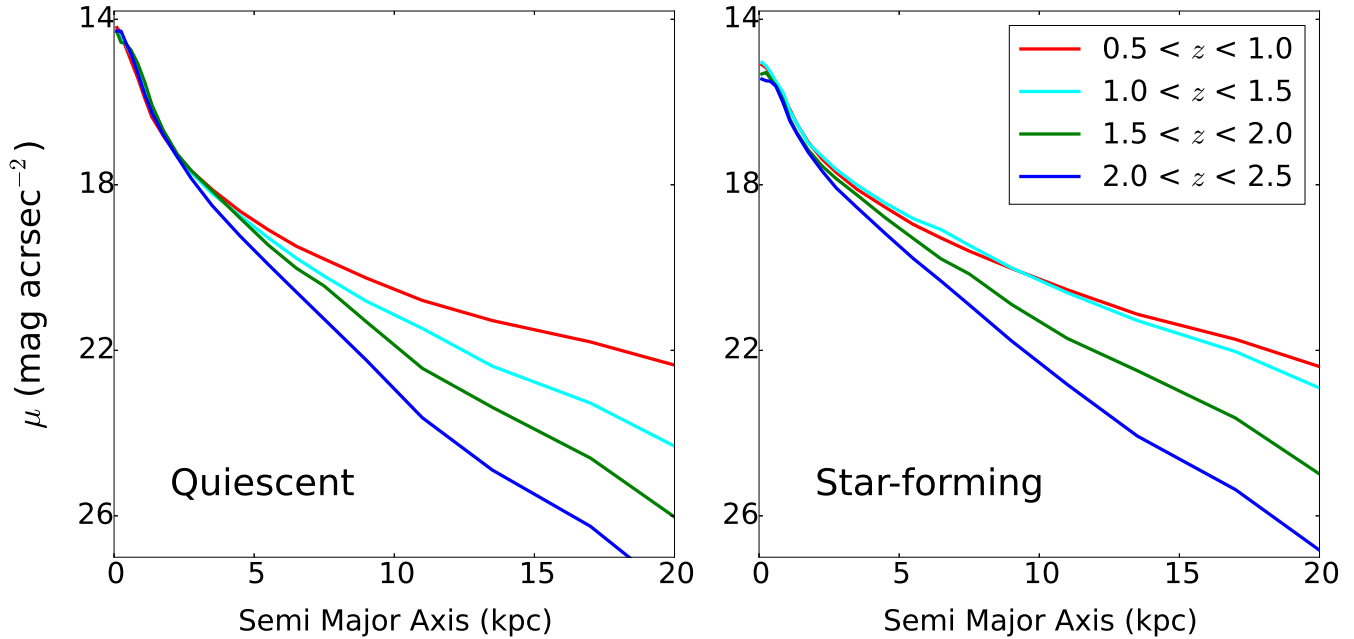}
  \caption{The inside-out growth of massive galaxies. Median light
    distributions of massive quiescent and star-forming galaxies are
    shown in four redshift bins. While the inner few kpc of these
    galaxies has been almost intact since \emph{z} $\approx$ 2.5,
    over time more material is accreted in their outskirts.
    Accretion onto quiescent galaxies continues at least down to \emph{z}
    $\approx$ 0.5, while it seems that star-forming galaxies stop accreting by
    $\emph{z}$ $\approx$ 1.0. The inner region of quiescent galaxies
    are brighter and have a higher density than the centers of star-forming
    galaxies. \label{fig:inside_out}}
\end{figure*}

\begin{figure*}[t]
  \centering
\includegraphics[width=180mm]{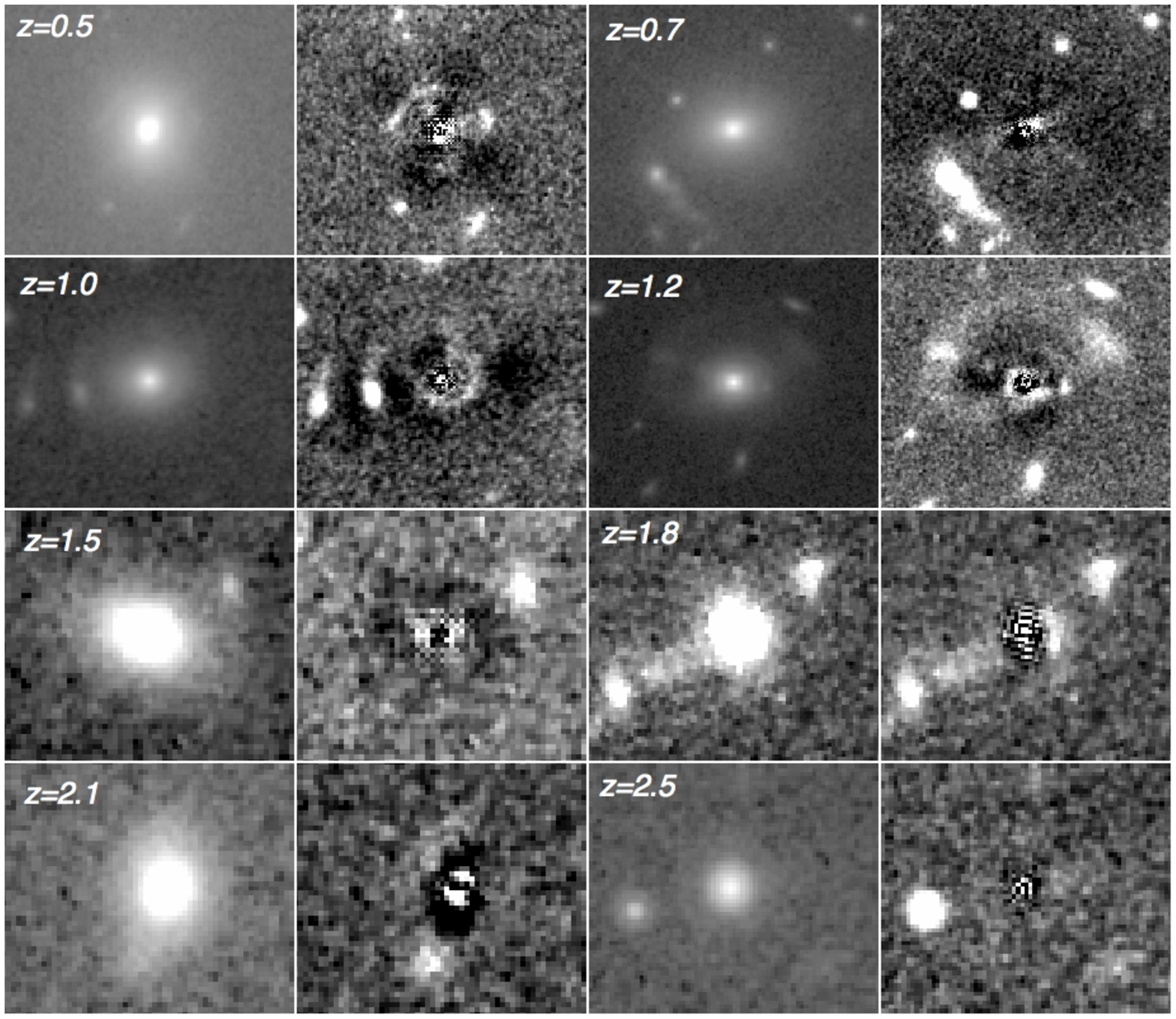}
  \caption{Examples of galaxies with tidal features or potentially nearby
    neighbors, at different redshifts.  The residual images,
    after removal of the bulge+disk model from the original galaxy
    image, allows for more effective visual detection of non-axisymmetric
    features.  \label{fig:merger_images}}
\end{figure*}

Davari et al. (2014, 2016) demonstrate that large S\'{e}rsic indices ($n > 6$) 
derived from single-component fits can lead to significant biases, which can 
be remedied by fixing $n$ to 6, after testing the fit for different initial 
guesses. This study follows the same general rule, except that the diagnostic plots are 
given more weight. For example, if fixing $n$ to 4 or 8 gives a better fit and 
cleaner residuals than fixing $n$ to 6, then those values are used. Another 
common symptom of unreliable fits is when the effective radius of the bulge 
drops below 0.5 pixel.  For many of these cases, changing the initial guesses 
of $R_e$ and $n$ leads to more realistic solutions.  But if the problem 
persists, we resort to fixing the bulge $R_e$ (or sometimes $n$ or both) to 
different initial values, and we rely on visual inspection of the residuals to
judge the merit of each model. For about 25\% of the
  galaxies at \emph{z} $>$ 1.5, the bulge $R_e$ (mainly to $R_e$=1) and/or bulge
  S\'{e}rsic index (mainly to $n$=1) are fixed.

In short, the goodness of single and two component
  fits are mainly  determined by visual inspection of the residual
  images (galaxy - model), along with the derived Sersic profile
  parameters. The derived Sersic parameters of each component have to
  be reasonable for a fit (e.g., $R_{e_{\rm bulge}}$/$R_{e_{\rm disk}}$ $<$ 1, ellipticities $<$
  0.8, sizes $>$ 0.5 pixel, and etc.) to be considered reliable. The
  objective of this study is not fitting two components only when
  there is an improvement over the single Sersic fit residual. For
  example, Figure 4 shows an example of a bulge-dominated galaxy,
  where adding a second component does not improve the residual
  significantly. However, the derived two component fit parameters not
  only confirm that this galaxy is bulge dominated (high
  $B/T$), but also provides additional information (e.g., $R_{e_{\rm bulge}}$,
  $n_{\rm bulge}$, $h$, and etc.)  Out of 248, only 1 and 7 galaxies could not be fit reliably with a
single S\'{e}rsic and two components, respectively, and are omitted from the
following analysis. In most of these 
cases, the image contains multiple regions with nonuniform and anomalous 
background values.

\section{Results}

\subsection{Fitting Galaxies with Single S\'{e}rsic Component}

Single S\'{e}rsic fitting is probably the most widely adopted method in the 
literature for morphological studies. This method provides a rather 
straightforward way for evaluating some key morphological properties of 
galaxies, namely size (usually parameterized as the effective radius; $R_e$), 
S\'{e}rsic index ($n$), and ellipticity ($e$). For instance, if a randomly 
distributed galaxy population has a significant disk component, we expect a 
wide distribution of $n$ and $e$.  The simulations of Davari et al. (2014, 
2015) show that single-component fits of massive galaxies at the redshift 
range of current interest ($0.5 < z < 2.5$) can be measured with little to no 
systematic uncertainty.

Figure \ref{fig:singleSersic} summarizes the results of single-component fits
of our sample, highlighting the redshift evolution of $R_e$, $n$, and $e$, 
separately for quiescent and star-forming galaxies.  For reference, we 
overplot the median value for a sample drawn from the second data release of 
GAMA (\citealt{GAMA}; \citealt{GAMA2}); the sample mass range corresponds to 
our number density selection (i.e. $M_{\star}=10^{11.1}-10^{11.3}\,M_{\odot}$) 
at $0 < z < 0.5$. The morphological parameters of the GAMA survey are derived 
by single S\'{e}rsic fitting, consistent with our method.  The error bars 
correspond to the interquartile range of different measurements. 

We perform two statistical tests to quantify the significance of the observed 
evolution of different properties: two-sample Kolmogorov-Smirnov (KS) test and 
one-way analysis of variance (ANOVA). The KS test is used to determine whether 
the star-forming and quiescent samples are drawn from the same parent 
population.  As a non-parametric test, it has the advantage of making no 
assumption about the distribution of data.  The results of KS test are 
summarized by the $D$-value and the $p$-value. The $D$-value shows the maximum 
difference between the empirical cumulative distribution functions of the two
samples, while the $p$-value indicates the significance of the difference
between two samples. Small $p$-values ($<0.05$) reject the null hypothesis 
that the two samples are drawn from the same distribution. ANOVA tests 
whether there are any statistically significant differences between the means 
of sample quantities in our redshift bins. The $F$-value (i.e. $F$ statistics) 
quantifies the variance between groups compared to the variance within groups: 
(variance of sample means at different redshift bins)/(variance of the whole 
sample). High $F$-values (i.e., small $p$-values) reject the null-hypothesis 
that the mean values are not statistically different at different redshift 
bins. In this study, high $F$-values indicate there is evolution over the 
observed redshift range.  The results of the KS and ANOVA test for the 
single-component fits are listed in Table 2.

Massive galaxies have experienced significant size evolution (top panel of 
Figure \ref{fig:singleSersic}; Table 2), with the size increase being more 
prominent for quiescent galaxies (Table 3).  Quiescent galaxies have increased 
their sizes by a factor of 3 down to $z = 0.5$, and more than a factor of 5 by 
$z \approx 0$.  By contrast, star-forming galaxies have undergone more modest
size growth, by a factor of $\sim 3$ down to $z = 0.5$.  However, the 
absolute amount of size increase  between $\emph{z}$ = 2.5 and $\emph{z}$ = 0.5 
for both star-forming and quiescent galaxies is comparable, about 2 kpc. The 
slope of the size-mass relation is consistent with the value found in 
\citet{vanDokkum10b} and \citet{Patel13}. On average, the size interquartile 
ranges are smaller for quiescent galaxies, which indicates a greater size 
diversity among star-forming galaxies. In other words, the sizes of quiescent 
galaxies are more homogeneous. Furthermore, the star-forming galaxies at each
redshift are larger than their quiescent counterparts, in agreement with
previous similar studies (e.g., \citealt{Zirm07}; \citealt{Szomoru11}; 
\citealt{Whitaker11}; \citealt{Patel13}; \citealt{Williams14}).  Star-forming 
and quiescent galaxies have statistically different size distributions 
(Table 2).

The global S\'{e}rsic indices of  both star-forming and quiescent galaxies in 
our highest redshift bin cluster around $n\approx 2.5$ (middle panel of Figure 
\ref{fig:singleSersic}), an intermediate value consistent with a composite 
bulge+disk system.  But over time, the two galaxy types diverge (Table 2).  
While star-forming galaxies maintain an almost constant $n$, the S\'ersic 
indices of quiescent galaxies increases significantly and systematically 
toward lower redshifts, eventually converging to resemble those of local 
elliptical galaxies ($n > 4$) at the lowest redshift bin. These trends are 
broadly consistent with the results of \citet{Morishita14}, \citet{Patel13}, 
and \citet{Szomoru11}.

The trends with regards to ellipticity are less definitive.  If massive 
galaxies initially host a sizable disk, we expect the eventual disappearance 
of that component to produce a notable reduction in the typical ellipticity 
of the population.  In practice, however, the presence of a sizable bulge 
concentration severely dilutes the expected ellipticity signature of any disk 
component.  Indeed, neither the ANOVA nor the KS test indicates any 
statistically significant redshift evolution of $e$.  (Table 2).  
However, considering only the quiescent galaxies at
  $z$ $>$ 1.5 and $z$ $<$ 1.5, there are two suggestive indicators of ellipticity evolutions. First,
F-test of equality of variances gives a p-value of 0.08 which is a
bordering signature of greater range (variance) of ellipticities at
$z$ $>$ 1.5. Second, comparing ellipticities of the quiescent galaxies at
$z$ $>$ 1.5 and $z$ $<$ 1.5 shows a tentative drop in the overall ellipticity: the two
sample one-sided t-test gives a p-value of ~0.05. These signatures are
suggestive and not conclusive. The following section presents a more
detailed analysis which provides an independent gauge for the presence
of a disk among massive quiescent galaxies at higher redshifts.

To summarize the results and implications of the single-component fits: the 
global light distribution of the massive galaxy population evolves 
significantly from $z = 2.5$ to $z = 0.5$.  Apart from the well-known increase 
in size, the population as a whole, and in particular the quiescent systems, 
exhibits systematical evolution toward larger S\'ersic indices and lower 
ellipticities at lower redshifts, converging to typical values of local 
ellipticals.  These trends support the thesis that the progenitors of 
present-day ellipticals were born with a sizable large-scale disk, which over 
time has been transformed.

\subsection{Fitting Galaxies with Two Components}

While single S\'{e}rsic fitting provides a reliable first-order estimate of 
morphological properties, decomposing a galaxy into its bulge and disk 
components can reveal a new set of valuable galaxy evolution indicators. The 
simulations of Davari et al. (2016) show that gross photometric properties,
in particular $B/T$, of bulge+disk systems usually can be measured accurately,
up to $z\approx 2-2.5$, without imposing any constraints on the profile shape 
of the bulge.  However, due to the inherent limitations of resolution, even 
with {\it HST}, detailed properties of the bulges (e.g., $R_e$ or $n$) can be 
measured reliable only for galaxies with $B/T$ $\geq$ 0.2.  The disk component,
by contrast, can be measured with little difficulty.

Figure \ref{fig:BT} depicts the overall variation of $B/T$ with redshift for 
our sample.  At higher redshifts, quiescent galaxies have intermediate values 
of $B/T$ ($\sim$0.4), but over time they become more and more bulge-dominated.
At the lowest redshift bin, $B/T \approx 0.8$, very close to the 
median value of local massive elliptical galaxies.  
Although  star-forming galaxies, too, become more bulge-dominated with time, their $B/T$ 
at all redshift bins are lower than that of their quiescent counterparts; the 
two classes have statistically different distributions in $B/T$ (Table 4). 
\citet{Bruce12} report that massive galaxies at $z>2$ are mostly disk-dominated
and by $1 < z < 2$ have increased their $B/T$ to intermediate values, with 
very few elliptical-like galaxies down to $z=1$. They show that disk-dominated 
galaxies have higher star formation rates, which translates into star-forming 
galaxies having a lower $B/T$.  Similarly, \citet{Lang14} also find that 
massive galaxies increase in $B/T$ between $1.5<z<2.5$ and  $0.5<z<1.5$.

Although $H$-band images of galaxies at different redshifts capture the flux
in different rest-frame bands (i.e., approximately $V$ to $I$ band),
multi-wavelength studies of nearby galaxies find that $B/T$ does not strongly
depend on observed rest-frame wavelength, at least within the standard optical
bands (e.g., \citealt{Schulz03}; \citealt{Graham08}).  The observed variation
of $B/T$ between different bands is less than $\sim 0.1$.  The shallow color
gradients of quiescent galaxies (e.g., Wirth 1981) further minimizes the
impact of rest-frame wavelength on $B/T$.

While the bulges of both types of galaxies become more luminous over time, the 
disks component behaves markedly differently: it becomes sub-dominant in 
quiescent galaxies but brightens for the star-forming group (middle and bottom 
panels of Figure \ref{fig:BT}; Table 4). Meanwhile, the star-forming galaxies
disks are becoming brighter at lower redshift bins. The bulges of quiescent 
galaxies attain higher luminosities than in star-forming galaxies at all 
redshifts, and in the lowest redshift bin the luminosities of quiescent 
galaxies have significantly smaller scatter than in star-forming galaxies.  

Our bulge+disk decomposition (Figure \ref{fig:Bulge} and Table 3) reinforces
the size evolution observed in the single-component fits 
(Figure \ref{fig:singleSersic}). The effective radii of the bulges of both 
classes have grown, and, once more, the evolution is steeper for quiescent
galaxies. The disk scale lengths of both star-forming and quiescent galaxies 
have increased (Figure \ref{fig:Disk}) as well, but their distributions are not
distinguishable (Table 4). Table 3 indicates that the disk size increase is 
less significant compared to the bulge component.

As shown in Figure \ref{fig:Bulge}, the S\'{e}rsic indices of quiescent galaxy 
bulges have increased considerably but have stayed almost the same for the 
star-forming population. This is similar to the results of single-component 
analysis (Figure \ref{fig:singleSersic}). By redshift 0.5, bulges of quiescent 
galaxies have S\'{e}rsic indices similar to that of typical local ellipticals 
and classical bulges (\citealt{Fisher08}).  

The disk ellipticities of both classes have similar distribution and have 
not changed between \emph{z} = 2.5 and 0.5.  On the other hand, the bulges of 
massive galaxies have become rounder over this period. Quiescent galaxies have
lower bulge ellipticities, and by \emph{z} = 0.5, their distribution is 
similar to that of local massive ellipticals and classical bulges 
(\citealt{Fathi03}).

In related studies, Bruce et al. (2014a, 2014b)
  analyze the rest-frame optical morphologies of
  a mass-selected sample of massive ($M_{\star} > 10^{10.5}\, 
M_{\sun}$) galaxies at $1 < z <  3$ in the CANDELS UDS and COSMOS fields. Similar to
  our work, they decomposed $H_{160}$-band images of massive galaxies into
  their bulge and disk components. In general, our results are in
  agreement with those of these authors. Bruce et al. (2014a) find that from $z$
  = 3 to $z$ = 1 the galaxies transition from disk-dominated to more
  bulge-dominated (their Figure 6), in accordance with our findings
  (our Figure 6). The results of Bruce et al. (2014b) show that 
  bulges exhibit a stronger size evolution than disks
  (their Table 3), with star-forming galaxies having relatively
larger disk sizes compared to passive systems (their Figure 8),
in qualitative agreement with our results (compare with our Table 3 and Figure 8). 
However, with regards to the the bulge components, they show that star-forming galaxies have
larger bulges than quiescent galaxies, contrary to the results
from our study. This may be due to
the fact that the bulge-disk decomposition of 
  Bruce et al. (2014a, 2014b) was done by fixing the S\'ersic index of the
  bulge and disk components to 4 and 1, respectively. As demonstrated
  in Davari et al. (2016), fixing the S\'ersic indices can lead to
  biases and larger uncertainties in measuring bulge and disk
  properties, depending on their size, redshift, and $S/N$.  Furthermore,
assuming a fixed profile for disks and bulges at all redshifts
precludes  any investigation of the evolution of these parameters with
look-back time.

\subsection{Further Evidence of Prominent Stellar Disks: Detection of 
Spiral Structures and Edge-on Disks}

Despite the prevalence of spiral structures and bars in the local Universe 
(e.g., \citealt{Lintott11}; \citealt{Willett13}), these features are not 
believed to be common among star-forming galaxies at higher redshifts
(\emph{z} $>$ 1.5) (e.g., 
\citealt{Conselice05}; \citealt{Bournaud09}; \citealt{Conselice11}), where 
the disks may be too dynamically hot (\citealt{Genzel06}; \citealt{Law07}; 
\citealt{Law09}).  

Our model-subtracted residual images yield an unexpected surprise: a sizable 
fraction of the sources exhibit spiral structure (Figure 
\ref{fig:spiral_images}) \footnote{Some examples can also be seen in Figures C1 
and C3 of \citet{Bruce12}.} .  All the cases are star-forming galaxies; there 
are no quiescent galaxies with securely detectable spiral structure (Table 5).
The case with the highest redshift is at $z = 2.4$.  The fraction of 
star-forming galaxies with spiral structure is $\sim 20$\% at $1.5 < z < 2.0$, 
and by $1.0 < z < 1.5$, the spiral fraction reaches nearly 70\%.  The fraction 
of star-forming galaxies with spiral structures drops (to 30\%) at the lowest 
redshift bin. This decline may not be reliable for two reasons.  First, the 
star-forming sample size at $0.5 < z < 1.0$ is very small (only 10 objects), 
and therefore the sample proportions are not statistically significant. 
Second, the $H$-band images for the low-$z$ objects are missing the bluer 
parts of the galaxy flux (see also \citealt{Elmegreen14}).  

The conditions necessary for the formation of spiral arms are complex (see
\citealt{Dobbs14} for a review), but one requirement is clear---the existence 
of a disk.  Thus, from the point of view of one of the main themes of this 
paper, the clear detection of spiral features in high-$z$ massive galaxies 
constitutes arguably the strongest, most model-independent evidence for the 
presence of a substantial disk component in these systems.  We see the spiral
features only in the star-forming galaxies and not in quiescent systems, but, 
by analogy with local S0 and spiral galaxies, this is not surprising.

An edge-on view of a galaxy can reveal another indisputable signature of a
prominent stellar disk.  Figure \ref{fig:thinDisk} gives several examples of 
highly flattened ($e > 0.6$) galaxies in our sample that are consistent with 
disk structures seen edge-on.  Interestingly, most of them are relatively 
thick. The fraction of galaxies with a stellar disk can be infered
from the frequency of detected edge-on galaxies. For this estimation, we assume a uniform
distribution of ellipticity with $0 < e < 0.8$ for a population of
bulge+disk systems. The inferred fraction of galaxies with a disk
hovers around 40--50\% at $1.0<z<2.5$,  
both among star-forming and quiescent galaxies, but below $z=1$, the incidence 
of edge-on quiescent galaxies drops to zero (Table 5). Highly
flattened systems (especially at high-\emph{z}) can be hallmarks of
merger. However, considering the fact that the majority of these
galaxies are compact, the chance of this degeneracy is low.

\section{Implications for Galaxy Evolution}

The discovery of red nuggets has captured much attention in recent years as it
requires a new paradigm for the formation and evolution of massive elliptical
galaxies.  The observed compactness of red nuggets at $z \approx2$ initially 
raised the question of whether the massive red galaxies have indeed increased 
their sizes by a factor of  roughly $\sim 3-5$ while maintaining their passive 
state, or whether the size measurement might in some way be flawed.  Extensive
recent simulations (\citealt{Mosleh13}; Davari et al. 2014, 2016), coupled with 
consistent results from multiple independent studies (e.g., \citealt{Daddi05}; 
\citealt{Toft07}; \citealt{Trujillo07}; \citealt{Buitrago08}; 
\citealt{Cimatti08}; \citealt{Franx08}; \citealt{vanderWel08}; 
\citealt{vanDokkum08}; \citealt{Damjanov09}; \citealt{Cassata10}; 
\citealt{Newman12}; \citealt{Szomoru12}), have minimized skepticism on the 
fidelity of the size measurements.  

This work confirms that massive galaxies at $z \approx 2$ were indeed 
compact, and over the next 10 billion years their sizes have increased 
significantly (Figure \ref{fig:singleSersic}).  The growth occurred inside-out
(\citealt{Patel13}; \citealt{Huang13b}).  Figure \ref{fig:inside_out} shows 
the median light distribution of massive quiescent and star-forming galaxies 
in four redshift bins.  While the inner few kpc of these galaxies have been
in place since $z \approx 2.5$, over time more and more material was
added to their outskirts. Accretion onto quiescent galaxies continued at 
least down to $z \approx 0.5$, whereas their star-forming counterparts seem to 
have stopped growing by $z \approx 1$. 

The compactness of the red massive galaxies at higher redshifts and their 
similarities to SMGs implicate the importance of strong gas dissipation during 
their early formation epochs, which in turn led to the starburst activity and 
accompanying disk formation (e.g., \citealt{Targett13}; \citealt{Toft14}). 
This raises some important questions: Do red nuggets at $z \approx 2$ have a 
sizable disk component, and if so, how prevalent was it?  And since red 
nuggets are widely believed to evolve into present-day elllipticals---indeed, 
our number density selection was specifically chosen to ensure that they 
do---can we trace the redshift evolution of the morphological transformation 
that must take place?

Table 5 shows the estimated fraction of galaxies with a prominent stellar disk,
using three different diagnostics: galaxies with $B/T$ $<$ 0.5, visually 
detectable spiral features (Section 4.3), and inferences from the frequency of 
detected edge-on ($e > 0.6$; Section 4.3) systems.

Our results suggest that disks may be common among high-\emph{z} massive galaxies, although 
it is difficult to obtain a conclusive estimate of their frequency. They range from 
an absolute minimum of $\sim 5$\%, as deduced from the incidence of spiral 
arms among star-forming systems, to as high as $\sim 80$\% for all massive 
galaxies regardless of star formation activity, according to bulge-disk 
decomposition.  The three disk diagnostics are not
  equally reliable and informative. While, visually detectable spiral
  structures are the most reliable indicator of a disk, they provide
  the lower limit, as the viewing angle and surface brightness dimming
  can prevent the detection of these structures. Furthermore, at higher
  redshifts, disks of massive galaxies may not be favorable to 
  long lived spiral structures. On the other hand, low $B/T$ at higher
  redshifts does not necessarily mean a disk resides in a galaxy and
  provides an upper limit. The fitted exponential component is not
  necessarily an indicator of a disk. Lastly, the inferred fraction of disks
  using the fraction of edge-on disks is probably the best proxy for
  the detection of disks. 

The prevalence of large-scale disks at $z\approx 2$ is further 
reinforced by the moderate S\'ersic indices and broad distribution of 
ellipticities derived from the global light distribution, a trend already 
echoed in other recent investigations.  By tracking the population from 
$z \approx 2.5$ to 0.5 and performing a consistent analysis of the whole 
sample, we witness the gradual transition of the large-scale morphology.  By 
$z \approx 0.5$, the red massive galaxies attain a high bulge fraction of 
$B/T \approx 0.8$, signifying the near disappearance of a dominant disk 
(Figure \ref{fig:BT}); their global (Figure \ref{fig:singleSersic} and bulge 
(Figure \ref{fig:Bulge}) S\'{e}rsic indices converge to $n \approx 4-5.5$, 
values close to that of de Vaucouleurs' profile; and their global axial ratios
drop to values closely resembling those of local massive galaxies (e.g., as 
measured in the GAMA survey).   All of these indicators strongly support the 
thesis that high-$z$ red nuggets are, in fact, the direct ancestors of today's
massive ellipticals.  

Much attention has been focused recently on the pivotal role that minor, dry 
mergers play in the evolution of red nuggets into present-day elliptical 
galaxies.  Minor mergers are considered the most plausible mechanism for 
explaining the dramatic size growth of massive quiescent galaxies (e.g., 
\citealt{Bournaud07}; \citealt{Bezanson09}; \citealt{Hopkins09}; 
\citealt{Naab09}; \citealt{vanDokkum10b}; \citealt{Oser12}; \citealt{Hilz13}),
their mass increase after being quenched (\citealt{vanDokkum10b}), the 
multiple-component structure of local ellipticals (Huang et al. 2013a, 2013b), 
and the prevalence of tidal features seen in deep imaging of nearby massive 
galaxies (\citealt{vanDokkum05}; \citealt{Tal09}; \citealt{Janowiecki10}).

The simulations of \citet{Welker15} stress the effectiveness of dry mergers in 
increasing the sizes of massive compact galaxies.  Consistent with 
\citet{Hilz13}, they find that dry mergers lead to a size-mass relation of 
the form $R_e$ $\propto$ $M^{\gamma}$, with $\gamma$ $\approx$ 2.  This is 
close to the size evolution we measure, $\alpha = -1.76\pm0.16$, for our quiescent massive 
galaxies between redshift 2.5 and 0.5.  Interestingly, star-forming galaxies 
have a considerably smaller value of $\alpha = -1.17\pm0.30$.  By $z\approx1$, 
more than 80\% of the simulated massive galaxies ($M_{\star} > 10^{10.5}\, 
M_{\sun}$) from \citet{Welker15} have experienced minor mergers. The merger 
rate is expected, on average, to increase monotonically with stellar mass 
(\citet{Hopkins10}), and therefore should be even higher for our sample. 
Figure \ref{fig:merger_images} shows a number of galaxies from our sample with 
disturbed morphologies and small-scale structure that may be indicative of 
merging activity.  We estimate, from visual inspection, that at $0.5<z<1.0$
more than 60\% of quiescent galaxies have small nearby objects or show merger
signatures (e.g., distortions, tidal tails, and shells); about 40\% of 
star-forming galaxies show similar features.  By $1.0 < z < 1.5$, the fraction
of merger candidates, for both classes combined, drops to $\sim 30$\%, 
presumably because it becomes increasingly difficult to resolve small-scale 
structure for distant galaxies.  While these morphological indicators are 
by no means secure estimators of the merger fraction, they at least give the
qualitative impression that mergers---especially minor mergers---play a part 
in the morphological transformation of massive galaxies.

The discovery of a significant disk component in massive galaxies at 
$z \approx 2$ and their eventual disappearance toward lower redshifts 
brings an important new element to the story.  How were the disks destroyed?  
Can this be accomplished by minor mergers alone?  Most likely not.  Breaking 
a big disk requires hitting it with something hefty, which can only be 
accomplished with a major merger.  The simulations of \citet{Hopkins10} show 
that major mergers are needed for forming galaxies with high $B/T$. 

Another key player in the evolutionary scenario of massive galaxies is, one 
that has captured less attention, are the compact blue galaxies
Although star-forming galaxies are larger than quiescent galaxies at each redshift bin, at high
redshifts these galaxies are compact, as well (Figure \ref{fig:singleSersic}). 
As local star-forming massive galaxies are rare (e.g., \citealt{Baldry12}), 
most of high-redshift compact blue galaxies must have also evolved into present-day 
massive ellipticals.  Figure \ref{fig:class_fraction} illustrates how over 
time the star-forming massive galaxies turn into quiescent massive galaxies. 
At the same time that the blue population quenches star formation, significant
morphological transformation must also occur to elevate the relative bulge 
fraction (Figure \ref{fig:BT}) and increase the S\'ersic index (both globally 
and for the bulge alone; Figure \ref{fig:singleSersic} and \ref{fig:Bulge}).

The prevalence of prominent stellar disks at higher redshifts raises the
possibility that some of these bulge+disk massive galaxies may have survived 
to the present.  Where are they?  The ``superluminous'' spiral galaxies 
discussed by Ogle et al. (2016) seem to fit the description. Ogle et al. quote 
an average number density of 32 Gpc$^{-3}$ at $z < 0.3$.  Interestingly, the
fraction of star-forming massive galaxies with spiral arms is 30\% at 
$z \approx 0.5$ (Table 5). As our overall sample was chosen to satisfy 
$n_c = 1.4\times10^{-4}$ Mpc$^{-3}$, the observed number density of massive 
spirals in our lowest redshift bin is $0.3 n_c$, or 42 Gpc$^{-3}$, very 
close to the average number density given by Ogle et al. 2015. (We note, however, that the
  sample size of star-forming massive galaxies at the lowest redshift
  bin is not statistically significant, as discussed in
  Section 4.3.)
\citet{Wellons15a}, using the Illustris (\citealt{Illustrisa};
\citealt{Illustrisb}) cosmological hydrodynamical simulations, trace the 
evolution of 35 massive compact galaxies from \emph{z} = 2. They find that 
$\sim$30\% of their galaxies survive undisturbed, while the rest have either 
experienced inside-out growth or have been destroyed via major mergers.

\section{Summary}

The discovery of massive compact galaxies at high redshift, specially red 
nuggets, has offered new insights into galaxy formation and evolution. These 
massive galaxies have major differences with their local counterparts, the 
massive ellipticals. They are not only compact but, as demonstrated in this 
study, they also possess a stellar disk. To match the population of 
present-day ellipticals, red nuggets must increase significantly in size 
{\it and}\ destroy their disks. Using a homogeneous and unbiased sample of $\sim$ 250 massive galaxies in the CANDELS
fields, spanning the redshift range 0.5 $<$ $\emph{z}$ $<$ 2.5, and selected through the fixed 
number density technique, we studied the evolution of morphological 
parameters as a function of redshift.  Further, we classified galaxies into quiescent
and star-forming systems using the UVJ color-color diagram in order to 
trace separately their evolutionary histories.

We conclude: 

\begin{itemize}

\item  The fraction of quiescent massive galaxies is higher at lower redshifts.

\item Both star-forming and quiescent galaxies have increased their
sizes significantly from $\emph{z}$ $\approx$ 2.5 to the present time, 
and the growth has occurred inside-out. 

\item The global S\'{e}rsic index of quiescent galaxies has increased over
  time (from $n$ $\approx$ 2.5 to $n$ $>$ 4), while that of
  star-forming galaxies has remained roughly constant ($n$ $\approx$ 2.5).

\item The distribution of global ellipticities has changed mildly with
  time, becoming rounder toward lower redshifts. 

\item The typical value of $B/T$ has increased with decreasing
  redshift, both for the quiescent and star-forming subsamples. By $z$
  ≈ 0.5, massive quiescent galaxies (with $B/T$ ≈ 0.8) begin to
  resemble the local elliptical galaxies. Star-forming galaxies have a
  lower median $B/T$ at each redshift bin. 

\item The  evolution of Sersic index, ellipticity, and $B/T$ suggests
  that both star-forming and quiescent galaxies have a significant
  stellar disk at early times, which systematically became less
  prominent toward lower redshifts. 

\item A considerable fraction of our sample have visually detectable
  spiral structures or thin disks observed nearly edge-on, which further confirms that
  high-\emph{z} massive galaxies have prominent stellar disks.

\item While minor dry mergers can explain the inside-out growth of massive galaxies, 
major mergers are needed to destroy their stellar disks between redshift 2.5 and the present time.

\item While the disks of star-forming and quiescent galaxies 
evolve similarly, their bulges follow
different evolutionary trajectories. The size increase of 
the bulges of quiescent galaxies is more
significant and their S\'{e}rsic indices and axial ratios are, on average, 
higher than their star-forming counterparts.

\end{itemize}

\acknowledgements
RD has been funded by a graduate student fellowship awarded by Carnegie Observatories. LCH acknowledges support by the Chinese Academy of Science through grant No. XDB09030102 (Emergence of Cosmological Structures) from the Strategic Priority Research Program and by the National Natural Science Foundation of China through grant No. 11473002.  RD thanks Heather Worthington-Davari for providing long-term support. 



\end{document}